\documentclass[journal]{IEEEtran}
\usepackage{amsmath,amsfonts}
\usepackage{algorithm}
\usepackage{algorithmicx}
\usepackage{array}
\usepackage{textcomp}
\usepackage{stfloats}
\usepackage{url}
\usepackage{verbatim}
\usepackage{graphicx}
\usepackage{cite}
\usepackage{booktabs}
\usepackage{multirow}
\usepackage{subfigure}
\def\BibTeX{{\rm B\kern-.05em{\sc i\kern-.025em b}\kern-.08em
    T\kern-.1667em\lower.7ex\hbox{E}\kern-.125emX}}
\usepackage{balance}

\begin{document}
\title{DiffCL: A Diffusion-Based Contrastive Learning Framework with Semantic Alignment for Multimodal Recommendations}
\author{Qiya Song, Jiajun Hu, Lin Xiao, Bin Sun~\IEEEmembership{Member,~IEEE}, Xieping Gao, Shutao Li~\IEEEmembership{Fellow,~IEEE}
}

\maketitle

\begin{abstract}
Multimodal recommendation systems integrate diverse multimodal information into the feature representations of both items and users, thereby enabling a more comprehensive modeling of user preferences. However, existing methods are hindered by data sparsity and the inherent noise within multimodal data, which impedes the accurate capture of users' interest preferences. Additionally, discrepancies in the semantic representations of items across different modalities can adversely impact the prediction accuracy of recommendation models. To address these challenges, we introduce a novel diffusion-based contrastive learning framework (DiffCL) for multimodal recommendation. DiffCL employs a diffusion model to generate contrastive views that effectively mitigate the impact of noise during the contrastive learning phase. Furthermore, it improves semantic consistency across modalities by aligning distinct visual and textual semantic information through stable ID embeddings. Finally, the introduction of the Item-Item Graph enhances multimodal feature representations, thereby alleviating the adverse effects of data sparsity on the overall system performance. We conduct extensive experiments on three public datasets, and the results demonstrate the superiority and effectiveness of the DiffCL. 
\end{abstract}


\section{Introduction}
\IEEEPARstart{R}{ecommender} systems (RSs), widely applied across various online platforms \cite{music2, online, Chen_Zou_Zhou}, represent a category of information filtering technologies designed to anticipate user preferences for various items and subsequently deliver tailored recommendations. These systems analyze users' historical behavioral patterns alongside relevant information to identify items or content that may align with their interests. Initially, early RSs predominantly relied on user interaction data, leveraging historical interaction to reflect behavioral similarities among users. However, as user needs become increasingly complex and diverse, these traditional systems have begun to reveal their limitations. In response to this challenge, multimodal recommender systems (MRSs) have emerged, emphasizing the integration of interaction information between users and items while incorporating diverse multimodal features. This approach enables MRSs to capture user preferences in a more comprehensive manner. MRSs have demonstrated considerable success in domains such as e-commerce and short video recommendations, where the richness of multimodal information enhances their ability to deliver personalized recommendations effectively. 
\begin{figure}[t]	
	\centering
	\subfigure[]{\includegraphics[width=0.98\columnwidth]{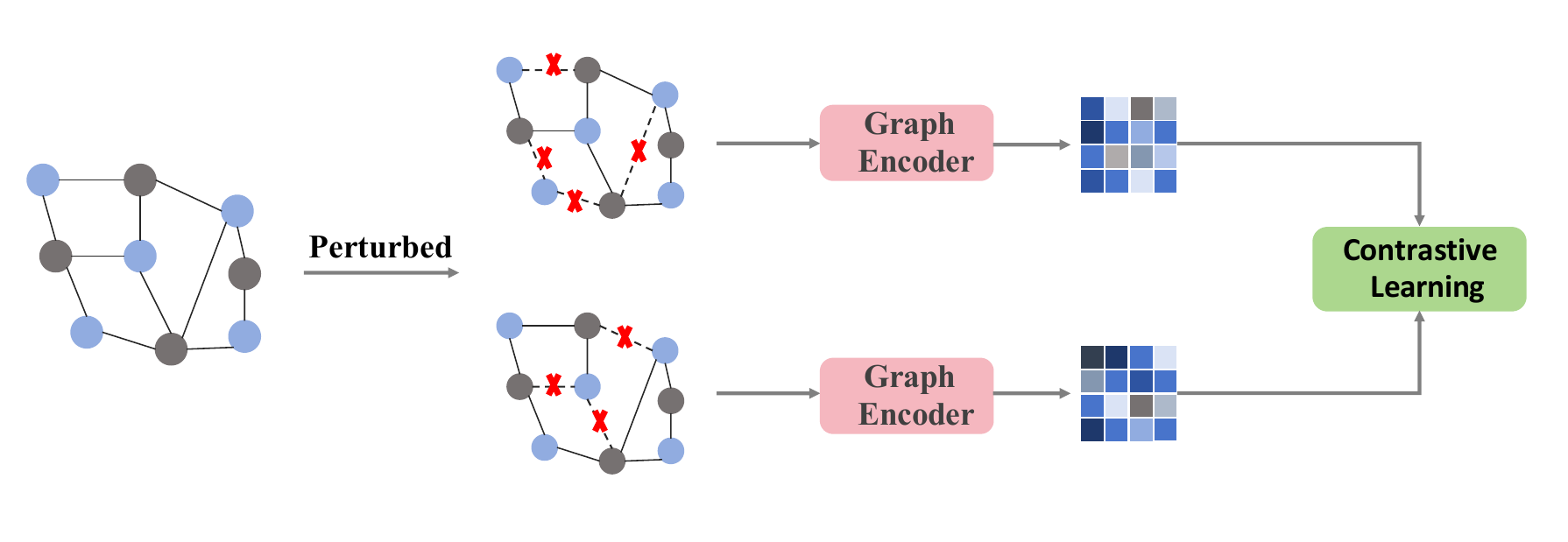}}
	\subfigure[]{\includegraphics[width=0.98\columnwidth]{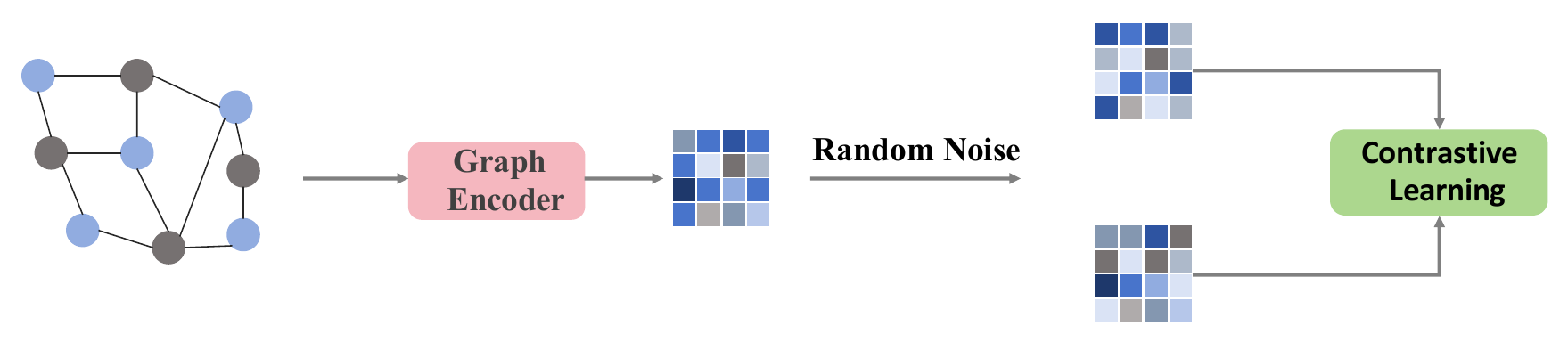}}
	\caption {Two methods for constructing graph contrastive learning: edge dropout and add random noise. (a) Some edges are randomly selected and removed from the graph according to a predefined dropout rate. (b) Add random uniform noise or gaussian noise to the feature embeddings after they have been processed by the Graph Encoder.}
	\label{1}
\end{figure}

Building upon conventional recommendation methods \cite{he2017neural,rendle2010factorization}, current MRSs employ diverse strategies to integrate information from multiple modalities. For instance, VBPR, as an extension of matrix factorization techniques \cite{rendle2010factorization}, introduces visual modal information into the feature representation of items for the first time, thereby effectively addressing the data sparsity challenge in RSs. Nevertheless, early methods primarily concentrated on unimodal data, resulting in models that are incapable of comprehensively modeling user preferences and item representations. As deep learning technology \cite{song2023multi,9755926,8736041} advances and gain widespread application, RSs leveraging deep learning have attracted significant attention from researchers. These methods \cite{wang2019neural, lei2020social} are capable of learning the underlying features of raw data and complex nonlinear correlations between items and users, thereby effectively modeling users' complex preferences. ACNE \cite{9451542} is an deep learning approach and its enhanced version (ACNE-ST) for modeling overlapping communities in network embedding, demonstrating superior performance in vertex classification task. CRL \cite{CRL} is based on matrix factorization, simultaneously considering the complementarity of global and local information, and collaboratively learning both topics and network embeddings.
Simultaneously, deep learning technology can represent different modal features in the same embedding space, which provides conditions for the rapid development of MRSs. To capture higher-order features of user interaction with the items, Graph Neural Networks (GNN) are utilized in RSs \cite{qiao2022tag, yu2020enhancing, yu2023xsimgcl}. It updates the representation of the current node by accumulating information from adjacent nodes within the graph, thereby integrating modal information through the message-passing process to yield a more comprehensive item representation. Existing GNN-based recommendation methods \cite{wei2019mmgcn,yu2023multi,wang2021dualgnn} require a significant amount of high-quality interaction data for modeling user-item relationships. Nonetheless, in practical applications of recommendation systems, interaction data is frequently sparse, which constrains the ability of recommendation models to produce accurate results.

Recently, self-supervised learning (SSL) generates supervision signals from unlabeled data, providing a new method for addressing the data sparsity issue in RSs \cite{yu2023self}. For instance, NCL \cite{lin2022improving} and HCCF \cite{xia2022hypergraph} combine SSL with collaborative filtering to model user-item interactions, yet they do not adapt them to specific multimodal recommendation scenarios. On the other hand, SGL \cite{wu2021self} employs dropout techniques to randomly eliminate interaction edges, subsequently constructing contrastive views to enhance item representations through contrastive learning. However, these methods concentrate exclusively on the enhancement of interactive data, neglecting the importance of considering multimodal information during the data augmentation process, which limits the model's capacity to representation multimodal information. Recent research has sought to address this gap by proposing the integration of multimodal modeling with self-supervised techniques \cite{yu2021self,zhou2023bootstrap} to enhance the accuracy of RSs. MMGCL \cite{yi2022multi} and SLMRec \cite{yu2021self} construct contrastive views for contrastive learning by perturbing the final modal features with added random noise. MMSSL \cite{wei2023multi} enriches the multimodal feature embeddings of items based on SSL with interaction information to achieve better recommendation results. Nevertheless, as shown in the Figure \ref{1}, these methods are generally based on intuitive cross-view contrastive learning or simple random augmentation techniques, which can introduce noise information irrelevant to the recommendation results to some extent.

Drawing inspiration from the success of diffusion models in the field of data generation, we propose a novel framework based on diffusion models for MRSs. Specifically, we first utilize graph convolutional networks to process pre-extracted raw multimodal features in order to capture higher-order multimodal feature representations. Subsequently, to construct a contrastive view that distinct from previous methods, we introduce diffusion models into the graph contrastive learning phase, utilizing both the forward and inverse processes of diffusion models to generate contrastive views, as opposed to merely adding random noise or employing dropout techniques. This approach effectively mitigates the impact of noise introduced during self-supervised learning. Finally, an Item-Item graph is introduced to augment the embedding of items, addressing the impact of data sparsity in RSs. Alongside this, we implement an ID-guided semantic alignment task to align semantic information across different modalities, enhancing semantic consistency. This alignment, guided by ID features, leverages their stability and uniqueness to ensure that the semantics of the item remain consistent, irrespective of the modality perspective. The contributions of this paper are briefly summarized as follows: 
	\begin{itemize}
	\item We propose a novel Diffusion-Based Contrastive Learning  framework (DiffCL) for multi-modal recommendation, which enhances the semantic representation of items by introducing Item-Item graphs to mitigate the effects of data sparsity.
	\item We introduce diffusion model to generate contrastive views during the graph contrastive learning phase, reducing the impact of noisy information in graph contrastive learning tasks.
        \item We utilize stable ID embeddings to guide the semantic alignment for enhancing consistency different modalities, thereby enabling effective complementary learning between the visual and textual modalities.
	\item We conduct extensive experiments on three public datasets to validate the superiority and effectiveness of the DiffCL.
	\end{itemize}
	
\section{Related Work}
\subsection{Multimodal Recommendation}
\noindent The multimodal recommendation system introduces multiple data sources as auxiliary information, extracts the semantic information corresponding to these auxiliary sources, and integrates them through multimodal fusion technology to obtain the user's multimodal preferences and the item's multimodal representation, which are used in final recommendation stage to improve recommendation accuracy. Early research focuses on single modality. For example, DUIF \cite{geng2015learning} builds on this foundation by utilizing additional user information to further enhance the user's feature representation. ACF \cite{chen2017attentive} utilizes an attention network to adaptively learn the weights of user preferences for items. In realistic scenarios, items exist with multiple modal information; therefore, utilizing different modal information at the same time allows for better modeling of user preferences. CKE \cite{zhang2016collaborative} is designed to enrich the feature representation of an item by combining image features and text features of item using a knowledge graph based on the matrix factorization technique. Wei et al. \cite{wei2019mmgcn} use multiple graph convolutional network to process different modal information as a way to extract cues about user's preference for particular modality. Zhang et al. \cite{zhang2021mining} provide a more comprehensive representation of items by building item semantic graphs to present hidden relationships between items. In this study, we endeavor to comprehensively leverage the multimodal features of items to model user interest preferences. To this end, we enhance item feature representation through the construction of an Item-Item graph, which uncovers latent relationships among items and establishes a more robust foundation for recommendations.
\subsection{Graph-based Models for Recommendation}
\noindent Graph Convolutional Networks (GCNs) have unique advantages in processing graph data structures, aggregate information from neighboring nodes, and facilitate the extraction of higher-order features \cite{GNN1,GNN2}. They are widely used in recommender systems. NGCF \cite{wang2019neural} fuses the GCN architecture based on matrix factorization, pursuing explicit encoding of higher-order collaborative signals to improve the performance of RSs. Several studies suggest that the nonlinear structures in GCNs are ineffective for extracting collaborative signals between users and items. Based on this, a lightweight GCN recommendation framework, LightGCN is proposed, which removes the original weight matrix and nonlinear activation function in GCN and achieves better recommendation effects. The GCN captures prevalence features other than collaborative signals, and the JMPGCF \cite{liu2022joint}, to match the user's sensitivity on prevalence, utilizes graph Laplace paradigm to capture prevalence features at multi-grained simultaneously. Recently, some contrastive learning methods are introduced into GCN-based recommender systems. SGL \cite{wu2021self} generates different contrast views through various dropout operations to perform contrastive learning based on them. MMSSL \cite{wei2023multi} introduces an inter-modal contrast learning task to retain semantic commonality across modalities, reducing the impact of noisy information on recommendation results. Building on these approaches, this work leverages graph convolutional networks to model user preferences and item multimodal features effectively. Furthermore, by incorporating a diffusion model into the graph contrastive learning phase, we aim to reduce the influence of multimodal noise, thereby achieving more robust and accurate recommendation outcomes.
\subsection{Diffusion Models for Recommendation}
\noindent In recent years, inspired by the diffusion model's (DM) wide application \cite{croitoru2023diffusion, ho2020denoising, lam2021bilateral} in the generative domain, some research combines DM with recommender systems to seek better recommendation results. For example, PDRec \cite{ma2024plug} leverages diffusion-based user preferences to improve the performance of sequential recommendation models. DreamRec \cite{yang2024generate} introduces DM to explore the latent connections of items in the item space and uses the user's sequential behavior as a guide to generate the final recommendations. DiffRec \cite{wang2023diffusion} uses DM in the denoising process to generate collaborative information that is similar to the global but personalized. Unlike the application of DM in the continuous item embedding space, LD4MRec \cite{yu2023ld4mrec} uses DM on the discrete item index and combines it with multimodal sequential information to guide the prediction of recommendation results. DiffMM \cite{jiang2024diffmm} enhances the user's representation by combining cross-modal contrast learning with a modal awareness graph diffusion model to better model collaborative signals as well as align multimodal feature information. In this work, we propose a novel graph contrastive learning component grounded in the diffusion model. Unlike conventional methods that rely on adding random noise to generate contrastive views, our approach harnesses the strong generative capabilities of DM to construct more informative and meaningful contrastive views, thereby enhancing the effectiveness of the learning process.

\section{Preliminary}
\noindent In this section, we present some preliminary knowledge related to the paper.
\subsection{Graph Neural Networks (GNNs)}
\noindent GNNs \cite{kipf2016semi} used to learn relationships between data from graph data structures. Where nodes denote entities and edges signify the relationships between these entities. The abstract structure of a GNN can be understood in terms of: message passing, aggregation and update functions. 

The message passing from neighbor $j$ to $k$ is represented by Equation \ref{gnn}. Each node $k$ collects information from its neighbors. The formula is as follows:
\begin{equation}
	m_{k}^{(l)} = \sum_{j \in \mathcal{N}(k)} M^{(l)}(h_{j}^{(l-1)}, h_{k}^{(l-1)}, e_{jk}),
	\label{gnn}
\end{equation}
here, $m_{k}^{(l)}$ is the aggregated message for node $k$ at layer $l$, $\mathcal{N}(k)$ denotes the set of neighbors of $k$, $h_{j}^{(l-1)}$ and $h_{k}^{(l-1)}$ are the feature vectors of nodes $j$ and $k$ from the before layer, and $e_{jk}$ denotes the edge features. The accumulation function combines these messages to form a new representation for the node: $h_{k}^{(l)} = U^{(l)}(h_{k}^{(l-1)}, m_{k}^{(l)})$, where $U^{(l)}$ is an update function that can take various forms, such as summation, mean, or a learnable neural network.
The foundational structure of Graph Neural Networks (GNNs) centers on the interplay of message passing, aggregation, and updating mechanisms, facilitating effective learning from complex relational data. This framework is highly versatile and can be adapted to address a wide range of tasks.
\subsection{Multimodal Fusion}
\noindent Multimodal fusion \cite{baltruvsaitis2018multimodal} is a powerful approach that integrates information from various modalities or sources to enhance the performance of machine learning models. By utilizing multiple types of data, multimodal fusion allows the system to offer a more holistic perspective of the underlying information. The process of multimodal fusion is divided into three main phases: feature extraction, alignment, and combination. During the feature extraction stage, pertinent features are identified and retrieved from each modality and converted into a unified representation. Alignment involves matching or correlating features across modalities to ensure that the information is contextually relevant. Finally, in the combination phase, the aligned features are merged using techniques such as early fusion, late fusion, or hybrid methods to generate a final representation that captures the strengths of each modality.
\subsection{Graph Contrastive Learning}
\noindent Graph contrastive learning is a self-supervised learning algorithm for graph data that optimizes graph embedding by constructing pairs of positive and negative samples. The core idea is to maximize the similarity between positive samples and the difference between negative samples to enhance the learning of graph representation. The most important part in graph contrastive learning is the graph augmentation strategy, which generates different views by randomly transforming (e.g. node dorpping, edge perturbation, attribute masking and subgraph) the original graph to increase the robustness and generalization ability of the model.

\section{Methodology}
\begin{figure*}[h]
	\centering
	\includegraphics[width=0.95\linewidth]{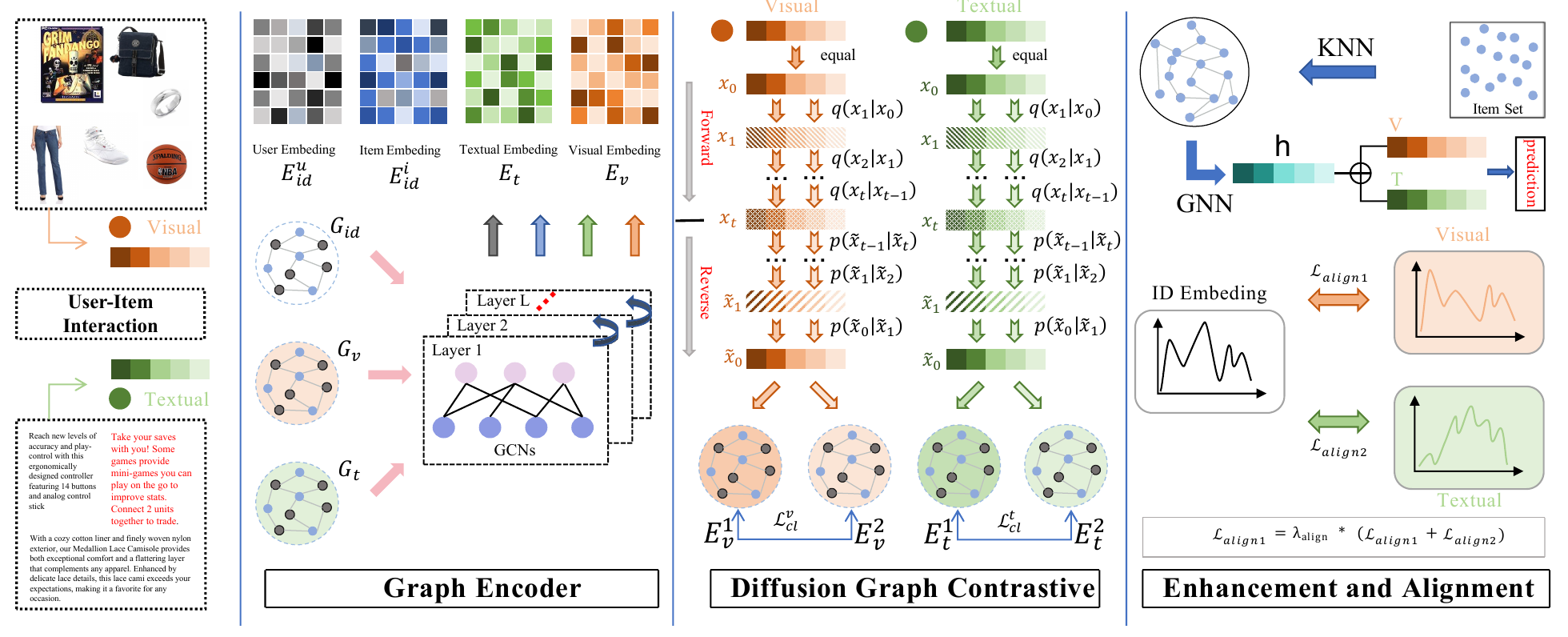}
	\vskip -0.15in
	\caption{The overview architecture of our  DiffCL framework. The DiffCL consists of three modules. \textbf{(a)} \textbf{Graph Encoder} is used to extract higher-order user preference cues and collaborative signals. The $G=\left \{G_{m}  \mid G_{v} , G_{t}, G_{id} \right \}$ represents three different user-item graphs. The \(E_{m} \) are the feature embeddings of different modalities obtained by GCNs, $m\in \left \{v, t, id  \right \} $. \textbf{(b)} \textbf{Diffusion Graph Contrastive} introduces diffusion model to construct contrast views for contrastive learning. \(E^{1}_{m}\) and \(E^{2}_{m}\) are feature embeddings obtained by diffusion model, $m\in \left \{v, t \right \} $. \textbf{(c)} \textbf{Enhancement and Alignment} implements item semantic enhancement and cross-modal semantic alignment. The ID embedding includes \(E^{u}_{id}\) and \(E^{i}_{id}\).}
	\label{structure}
	\vskip -0.1in
\end{figure*}

\noindent  In this section, we take an indepth exploration of the DiffCL framework. It includes what key components the DiffCL consists of and a mathematical description of those components. The detailed workflow of the DiffCL is illustrated in Figure \ref{structure}.

\subsection{Problem Formulation}
\noindent 
In the recommendation process, relying solely on user-item interaction data may lead to insufficient information, failing to fully reflect users' interests and preferences. Multimodal recommendation systems can capture users' needs more comprehensively by integrating data from different modalities, thus providing more personalized and accurate recommendations.

The process of multimodal recommender system is as follows, given $U=\left\{ u \right \}$ and $I=\left \{ i \right \}$ as the sets of users and items, respectively, the total counts are $|U|$ and $|I|$. We process the original user-item ID interaction to get the embedded feature $E_{id}$ of the ID, and then the visual modal features $E_{v}$ and textual modal features $E_{t}$ are obtained through different encoders. After a series of enhancement and alignment operations, we obtain the final item embedding $e_{i}$ and user embedding $e_{u}$. And the inner product of the two is calculated to obtain the predicted score $\hat{y}_{u,i}$ of user $u$ for the target item $i$, with the following formula:
\begin{equation}
	\hat{y}_{u,i} = e_{u} \cdot e_{i}^{T} .
	\label{yuce}
\end{equation}

The multimodal recommender system calculates the user's predicted scores for different items and orders them from highest to lowest score, taking the top-K items as the final recommended list.
\subsection{Graph Encoder}
\noindent
Some research on recommender systems based on GCNs has demonstrated that constructing item-user heterogeneous graphs and processing them with graph neural networks can better capture the user's preference cues, which in turn improves the recommendation accuracy of the whole system. Inspired by these works, we propose a graph encoder component that consists of three different GCNs to capture higher-order features of different modalities.

First, we use ResNet50 \cite{he2016deep} and BERT \cite{geng2015learning } to extract the image information and text information in the raw data, respectively, and encode the features of these modal information. After that, the graph encoder component is utilized to capture the higher-order features of user-item interactions, image and text features, respectively. 

Anchored in the interaction information in the raw data and the multi-model information of the items, we constructed three user-item graphs $G=\left \{G_{m}  \mid G_{v} , G_{t}, G_{id} \right \}$. We construct the interaction matrix $J$ to represent the interaction information, such that $j_{ui} = 1$ if an interaction exists between user $u$ and item $i$, otherwise $j_{ui} = 0$. $G_{m} =\left \{ n, e \right \} $ denotes user-item graph, $m\in \left \{v, t, id  \right \} $, $n$ represents the set of nodes and $e$ represents the set of edges in the graph. The features after the $l$-th layer of convolutional network are denoted as $E_{m}^{(l)}$ and the final embedded features are denoted as ${E}_m$ and they are mathematically expressed as follows:
\begin{equation}
	E_{m}^{(l)}=\sum_{i \in {N}_u} \frac{1}{\sqrt{\left|{N}_u\right|} \sqrt{\left|{N}_i\right|}} E_m^{(l-1)}, 
\end{equation}
where, $N_i$ is the single-hop neighbor of $i$ in graph $G_m$, $N_u$ is the single-hop neighbor of $u$ in graph $G_m$.
	\begin{equation}
	{E}_m = \sum_{l=0}^LE^{(l)}_m,
\end{equation}
where, $L$ is the layer count of the graph convolution, $E_{m}^{(0)}$ is the original feature after initial feature extraction.

\subsection{Diffusion Graph Contrastive Learning}
\noindent Over the last few years, diffusion model (DM) has excelled in the field of data generation, which can generate data that is highly consistent with the original data. Motivated by the application of DM, we propose a new multimodal recommender system method called DiffCL based on DM. Diffusion Graph Contrastive Learning is the most important component of the DiffCL. We introduce the DM to the graph contrastive learning phase and use it to generate two similar but inconsistent contrasting views to enhance the representation of items and users. Specifically, we gradually add Gaussian noise to the original user-item graph to destroy the original interaction information between the two, and then restore the original interaction by predicting the original state of the data through a probabilistic diffusion process.
\subsubsection{Graph Diffusion Forward Process}
The higher order feature captured by the Graph Encoder is $E_{m}$, which is represented by the following mathematical form:
    \begin{eqnarray}
    E_{m} & = & \begin{bmatrix} e_{m}^{u} & e_{m}^{i} \end{bmatrix},
    \end{eqnarray}
    where, \(m \in \left \{ v,t,id\right \} \), \(e_{m}^{u}\) and \(e_{m}^{i}\) denote user embedding and item embedding in a particular modality, respectively. 
    
    Our graph diffusion process includes only visual modalities and textual modalities, here we take visual modalities as an example to describe the diffusion process mathematically.
    We consider the visual modality of the embedding 
    \(E_{v} = \begin{bmatrix} e_{v}^{u} & e_{v}^{i} \end{bmatrix}\). We initialize the diffusion process with \(x _{0} = \begin{bmatrix} e_{v}^{u} & e_{v}^{i} \end{bmatrix} \). The forward process is a Markov chain that constructs the final \(\boldsymbol{x}_{1: T}\) by gradually adding Gaussian noise at each time step \(t\). Specifically, the process from \(x _{t-1}\) to \(x _{t}\) is expressed as follows:
    \begin{eqnarray}
    q\left(\boldsymbol{x}_{t} \mid \boldsymbol{x}_{t-1}\right) & = & \mathcal{N}\left(\boldsymbol{x}_{t} ; \sqrt{1-\beta_{t}} \boldsymbol {x}_{t-1}, \beta_{t} \boldsymbol{I}\right),
    \end{eqnarray}
    where, \(\mathcal{N}\) represents a Gaussian distribution, and \(\beta_{t}\) is a noise scale that regulates the increase of Gaussian noise at every time step \(t\). As \(t \to \infty \), \({x}_{t}\) will converge to a standard Gaussian distribution. Since independent Gaussian noise distributions are additive, it follows that we can get \({x}_{t}\)  directly from \({x}_{0}\) . This process is expressed by the formula: 
    \begin{eqnarray}
    q\left(\boldsymbol{x}_{t} \mid \boldsymbol{x}_{0}\right) & = & \mathcal{N}\left(\boldsymbol{x}_{t} ; \sqrt{\bar{\gamma}_{t}} \boldsymbol{x}_{0},\left(1-\bar{\gamma}_{t}\right) \boldsymbol{I}\right).
    \end{eqnarray}
    We utilize two parameters to control the total amount of noise added during the process from \({x}_{0}\) to \({x}_{t}\): \({\gamma_{t} }\) and \(\overline{\gamma_{t} } \). Their mathematical representation is as follows:
    
    \begin{equation}
       \gamma_{t} = 1 - \beta _{t} ,
    \end{equation}
        \begin{equation}
       \overline{\gamma_{t} } =\prod_{1}^{t}\gamma_{t}.
    \end{equation}
    Then \({x}_{t}\) can be re-parameterized as:
    \begin{equation}
        x_{t} =\sqrt{\overline{\gamma_{t} }} x_{0} +\sqrt{1-\overline{\gamma_{t}} }\varepsilon ,
    \end{equation}
    where, \(\varepsilon \sim \mathcal{N}\left (  0,\boldsymbol{I}\right ) \). We employ a linear noise scheduler for \(1-\overline\gamma\) to control the amount of noise in $x_{0:\textit{T}}$:
    
    \begin{eqnarray}
    1-\bar{\gamma}_{t} & = & s \cdot\left[\gamma_{\min }+\frac{t-1}{T-1}\left(\gamma_{\max }-\gamma_{\min }\right)\right]\ ,
    \end{eqnarray}
    where, \(t\in \left \{ 1,2,\cdots,T \right \} \), \(s\) is noise scale and \(s\in \left [ 0,1 \right ] \), \(\gamma_{\min }\) and \(\gamma_{\max }\) represent the maximum and minimum limits of additive noise.

\subsubsection{Graph Diffusion Reverse Process}
 \noindent Reverse process's objective is to eliminate the noise introduced through process from \({x}_{0}\) to \({x}_{t}\) and to recover \({x}_{0}\). This process generates a pseudo-feature similar to the original visual representation. The transformation associated with the reverse process begins at ${x}_{t}$ and gradually recovers ${x}_{0}$ through a denoising transformation step. The mathematical expression for the reverse process is as follows:
    \begin{equation}
        p_{\theta}\left(x_{t-1} \mid x_{t}\right)=\mathcal{N}\left(x_{t-1} ; \mu_{\theta}\left(x_{t}, t\right), \Sigma_{\theta}\left(x_{t}, t\right)\right),
    \end{equation}
\(\mu _{\theta}\left (x_{t} ,t \right ) \) and \(\Sigma_{\theta}\left(x_{t}, t\right)\) denote the predicted values of the mean and variance of the Gaussian distribution in the next state, respectively, and we obtain them by having two learnable parameter neural networks.

\subsubsection{Graph Contrastive Learning}
\noindent Our framework employs a widely used enhancement strategy that utilizes contrast learning to enhance modality-specific feature representations. Specifically, we utilize diffusion model to generate contrasting views. The visual representation of the graph contrastive learning of visual modalities, for example, is \(E_{v}\) after the graph convolution operation. Let \({x}_{0}=E_{v}\), we can get a representation \(E_{v}^{1} \) similar to \(E_{v}\) through diffusion model. Repeat the execution of the operation's to another representation \(E_{v}^{2} \). Then we perform graph contrastive learning based on InfoNCE \cite{he2020momentum} loss function with the following formula:
\begin{align}
\mathcal{L} _{u}^{v}=\sum_{u_{1} \in U} -\log_{}{\frac{\exp \left({s}\left({e}_{u_1, v}^{1} , e_{u_1, v}^{2}\right)\right) / \tau}
{\sum_{u_2\in U}^{}\exp \left({s}\left({e}_{u_2, v}^{1} , e_{u_2, v}^{2}\right)\right) / \tau } } ,\\
\mathcal{L} _{i}^{v}=\sum_{i_{1} \in I} -\log_{}{\frac{\exp \left({s}\left({e}_{i_1, v}^{1} , e_{i_1, v}^{2}\right)\right) / \tau}
{\sum_{i_2\in I}^{}\exp \left({s}\left({e}_{i_2, v}^{1} , e_{i_2, v}^{2}\right)\right) / \tau } }.
\end{align}

\begin{equation}
\mathcal{L}_{cl}^{v}=\mathcal{L} _{u}^{v}+\mathcal{L} _{i}^{v},
\end{equation}
here, $s\left ( \cdot  \right ) $ represents the cosine similarity function, and $\tau$ is a hyperparameter that controls the rate of model convergence. 

The same reasoning leads to:
\begin{equation}
\mathcal{L}_{cl}^{t}=\mathcal{L} _{u}^{t}+\mathcal{L} _{i}^{t}.
\end{equation}

The final loss of graph contrastive learning loss is as follows:
\begin{equation}
\mathcal{L}_{cl}=\lambda_{\mathrm{cl}}(\mathcal{L}_{cl}^{v} + \mathcal{L}_{cl}^{t}),
\end{equation}
where, $\lambda_{\mathrm{cl}}$ is a hyperparameter used to control the graph contrastive learning loss.

\subsection{Multimodal Feature Enhancement and Alignment}
	\subsubsection{Multimodal Feature Enhancement}
	\noindent To extract the semantic connections among various items, we build the Item-Item graph ( I-I graph), which is implemented by the KNN algorithm. Specifically, we calculate the similarity scores $S_{i,j}^{m} $ between different item pairs \( \left( i , j \right) \) separately based on different modal features to obtain I-I graphs for specific modalities. The similarity score is calculated by the following equation:
	
	\begin{equation}
		S^m_{i,j}=\frac{\left(f_i^m\right)^{\top} f_{j}^m}{\left\|f_i^m\right\|\left\| f_{j}^m\right\|},
	\end{equation}
	where, $i$ and $j$ denote pairs of items consisting of different items. $f_i^m$ and $f_j^m$ denote the original features of the specific modes of items $i$ and $j$, respectively. \(m \in \left \{ v,t\right \} \) represents the modality.
	
	 To reduce the effect of redundant data upon model accuracy, we selectively discard the obtained similarity scores. We keep only the K neighbors whose similarity scores are ranked in the top-K and assign them a value of $1$, which can be expressed as follows:
	\begin{equation}
		S^m_{i,j} = \begin{cases}1 & \text { if } S^m_{i,j} \in \text { top-} K\left(S^m_{i,j}\right) \\ 0 & \text { otherwise }\end{cases}.
	\end{equation}
	When $S^m_{i,j}$ is 1, it represents a potential connection between item pairs \( \left( i , j \right) \). Simultaneously, we fix $K$ = 10. We normalize $S^m$ utilize the following formula: 
	\begin{equation}
		\widehat{S}^{m}=\left(D^{m}\right)^{-\frac{1}{2}} {S}^{m}\left(D^{m}\right)^{-\frac{1}{2}},
	\end{equation}
	Here, $D^{m}$ represents diagonal matrix of ${S}^{m}$ and $D^{m} \in \mathbb{R}^{N\times N} $. 
	It generates a symmetric, normalized matrix that helps eliminate the influence of node degrees on the results, making subsequent aggregation operations more stable. The calculation formula for $D^{m}_{ii}$ is as follows:
	\begin{equation}
	D^{m}_{ii} = \sum_{j}{S}^{m}_{i,j}.
	\end{equation}
	Then, we aggregate multi-layer neighbor information based on the obtained modality-aware adjacency matrix:
	\begin{equation}
		A_{m}^{(l)} =\sum_{j\in N_{i} }^{} \widehat{S}_{i,j}^{m} A_{j_m}^{(l-1)},
	\end{equation}
	where, $j$ is first-order neighbor of $i$, $A_{j_m}^{}$ denote the embedding of item $j$ in modality $m$, \(m \in \left \{ v,t\right \} \).
	
	To better utilize various modal information to dig user preferences, we enhance the final embedding $E_m$ using the embeddings $A_{m}^{(l)}$ of the specific modal I-I graph, expressed by the formula:
	
	\begin{eqnarray}
		E_{m} & = & \begin{bmatrix} e_{m}^{u} & e_{m}^{i} + A_{m}^{(l)}\end{bmatrix}.
	\end{eqnarray}

	\subsubsection{Multimodal Feature fusion}
	\noindent Different modalities carry distinct modal information, which is both relevant and complementary. To more comprehensively capture user behavior preferences, we implement feature-level fusion for visual and textual features, resulting in the following fused feature representation:
	\begin{equation}
		{E}_{vt} = \mu  \times {E}_v + (1-\mu ) \times {E}_t,
	\end{equation}
	where, ${E}_v$ and ${E}_t$ denote the visual features and textual features, respectively, and $\mu $ is a trainable parameter with an initial value of 0.5.
	
	In the phase of feature fusion, we do not fuse the ID modality with other multimodal features. This is because, in multimodal recommendation systems, the ID modality possesses uniqueness and stability. Therefore, we only use it to align multimodal features and calculate the final predicted scores.
	\subsubsection{Multimodal Semantic Alignment}
	\noindent In multimodal recommendation systems, the feature distributions of different modalities are generally inconsistent, and the fusion process often retains a lot of noise information. Moreover, some existing modal semantic alignment methods disrupt historical interaction information, which adversely affects the final predictions. Therefore, we propose a cross-modal alignment method that uses stable ID features as guidance, effectively leveraging the ID embedding to better align semantic information of different modalities, ensuring semantic consistency among the information from various modalities. Inspired by the article PPMDR \cite{liu2023federated}, we parameterize the final ID modality feature $E_{id}$, the visual modality feature $E_v$, and the textual modality feature using a Gaussian distribution. Then, we calculate the distance between the ID modality and the visual modality, and the textual modality feature distributions as losses, respectively. The formula is as follows:
	\begin{equation}
		{E}_{id} \sim N\left(\mu_{id}, \sigma_{id}^2\right),
	\end{equation}
	\begin{equation}
	\left\{ 
	\begin{aligned}
		{E}_{v} \sim N\left(\mu_{v}, \sigma_{v}^2\right),  \\
		{E}_{t} \sim N\left(\mu_{t}, \sigma_{t}^2\right),
	\end{aligned}
	\right.
	\end{equation}
	\begin{equation}
		\left\{ 
		\begin{aligned}
			\mathcal{L}_{align_1} = |\mu_{id} - \mu_{v}| + |\sigma_{id} - \sigma_{v}|,  \\
			\mathcal{L}_{align_2} = |\mu_{id} - \mu_{t}| + |\sigma_{id} - \sigma_{t}|.
		\end{aligned}
		\right.
	\end{equation}
	The final alignment loss is calculated as follows:
	\begin{equation}
		\mathcal{L}_{align} = \lambda_{align}(\mathcal{L}_{align_{1}} + \mathcal{L}_{align_{2}}),
	\end{equation}
	where, $\lambda_{align}$ is hyper-parameter used to balance the alignment loss.

\subsection{Model Optimization}
	\noindent In recommendation tasks, Bayesian Personalized Ranking (BPR) is a commonly used optimization method. The basic idea of BPR is to increase the distinction between the expected scores of positive and negative samples, as it supposes users are more likely to prefer the items they have interacted with. We construct a triplet $(u, p, n) \in {D}$ for calculating the BPR loss, $u$ represents user, $p$ denote the items that have been interacted with by $u$ and $n$ denote the items that have been interacted with by $u$. The formula is as follows:
	\begin{equation}
		\mathcal{L}_{BPR} = \sum_{(u, p, n) \in {D}} - \log(\sigma(y_{u,p} - y_{u,n})),
	\end{equation}
	where, $y_{u,p}$  represents predicted score for $p$ by $u$, while $y_{u,n}$ denotes the predicted score for $n$ by the same user. Additionally, $\sigma$ refers to the sigmoid function.
	
		$y_{u,p}$ and $y_{u,n}$ are calculated by the following equations:
	
	\begin{equation}
		y_{u,p} = \left ( e_{vt}^{u} \right )^T \cdot e_{vt}^{p} + \left ( e_{id}^{u} \right )^T\cdot e_{id}^{p},
	\end{equation}
	
	\begin{equation}
		y_{u,n} = \left ( e_{vt}^{u} \right )^T \cdot e_{vt}^{n} + \left ( e_{id}^{u} \right )^T\cdot e_{id}^{n},
	\end{equation}
	where, $e_{vt}^{u}$ and $e_{vt}^{p}$ signify the representations of $u$ and $p$ following modal fusion, respectively. Meanwhile, $e_{id}^{u}$ and $e_{id}^{p}$ refer to the ID embeddings for $u$ an $p$.

	Finally, we combine BPR loss, diffusion graph contrastive learning loss, and cross-modal alignment loss to calculate the total loss, as shown in the following formula:
		\begin{equation}
		\mathcal{L}=\lambda_{\mathrm{cl}} \mathcal{L}_{\mathrm{cl}}+ \mathcal{L}_{align}+\mathcal{L}_{\mathrm{BPR}}+\mathcal{L}_{\mathrm{E}}
	\end{equation}
	$\mathcal{L}_{\mathrm{E}}$ is the regularization loss, the calculation formula is as follows:
	\begin{equation}
		\mathcal{L}_{E}=\lambda_{E}\left(\left\|E_{v}\right\|_{2}^{2}+\left\|E_{t}\right\|_{2}^{2}\right),
	\end{equation}
 	where, $\lambda_{E}$ is hyperparameter to regulate the impact of the $L_2$ regularization.

\begin{table}[ht]
	\centering
	\caption{Specific Data Distribution Of The Experimental Dataset}
	\scriptsize
	\setlength{\tabcolsep}{5mm}{
		\begin{tabular}{@{}lllllll@{}}
			\toprule
			Dataset & \#User & \#Item & \#Interaction & Spasity    \\ \midrule
			Baby    & 19,445 & 7,050  & 160,792    & 99.88\%  \\
			Sports  & 35,598 & 18,357 & 296,337    & 99.96\%  \\
			Video   & 24,303  & 10,672  & 231,780    & 99.91\%  \\ \bottomrule
	\end{tabular}}
	\label{data}
\end{table}
    
\section{Experiments}
\noindent To fairly appraise the performance of the DiffCL, we construct a large number of experiments to assess its performance. First, we compare the DiffCL with other state-of-the-art multimodal recommendation methods, this comparison is based on the same dataset processing. In addition, we construct different variants for the DiffCL to validate the effect of distinct components to ensure that we can effectively improve the final recommendation accuracy for each component. Finally, we set different hyperparameters for the model to search for the optimal hyperparameter settings. This experiment is set up to answer the following three questions: 
\begin{itemize}
	\item RQ1: How does the DiffCL's performance on different datasets compare to general RSs and MRSs?
	\item RQ2: How does the different components that make up the DiffCL affect its overall recommended performance?
	\item RQ3: How does setting different hyperparameters change the performance of the DiffCL?
\end{itemize}
\subsection{Experimental Settings}
\noindent In this paper, the dataset used is the Amazon review dataset, which is extensively used in MRSs. The initial dataset contains information about the user's interaction with the item, description of the text and images of the items, and the text of comments about the item from interacting users, as well as other information such as the price of the item. For all comparative models, we use the same way of processing the dataset. Specifically, we use 5-core filtering to filter the raw data and optimize the data quality. Prior to model training, we pre-extracted item's visual and textual features for the following recommendation task by utilizing the pre-trained ResNet50 \cite{he2016deep} and BERT \cite{devlin2018bert} with initial dimensions of 4096 and 384 dimensions for the visual and textual features, respectively. The data sets are as follows: (1) Baby, (2) Video, (3) Sports and Outdoors (denoted as Sports). Table \ref{data} shows the details of the three datasets. The training, validation and test sets are divided in the ratio of 8:1:1.

	\subsubsection{Baselines}
	In this subsection, we demonstrate the superiority of DiffCL by comparing it with current state-of-the-art recommendation methods. The comparative models used for the experiments include general recommendation model as well as multimodal recommendation model. Our purpose is use comparative experiments to demonstrate that DiffCL has some advantages over other models in the recommendation task. In addition, sufficient support for its application in real world is provided by experiments on real-world datasets.
	
		The general recommendation model and multimodal recommendation model used for the comparative experiments are as follows:
		
		\noindent (a) general recommendation methods:
		
		\textbf{BPR} \cite{rendle2012bpr} : This is a recommendation algorithm for implicit feedback that uses a Bayesian approach to model user preferences for items and randomly selects negative samples during training to improve model's generalization.
		
		\textbf{LightGCN}: It removes some unnecessary components from the graph convolutional network and improves the training efficiency of the model.
		
		\noindent (b) multimodal recommendation methods:
		
		\textbf{VBPR}: This method is based on an extension of BPR that introduces visual features of items to improve RS's performance for multimodal recommendation scenarios that include visual data.
		
		\textbf{MMGCN} \cite{wei2019mmgcn}: This method utilizes a graph structure to capture complicated relationships between users and items. It also designs specialized mechanisms to integrate information from various modalities, ensuring that information from various modalities can effectively complement .
		
		\textbf{DualGNN} \cite{wang2021dualgnn} : This method models the relationships between users and items simultaneously using GNN, capturing multi-level relational information to improve the accuracy and personalization of recommendations. 
		
		\textbf{SLMRec} \cite{tao2022self} : This method utilizes self-supervised learning by designing tasks to generate labels and adopts a contrastive learning strategy to optimize the model by constructing positive and negative sample pairs.
		
		\textbf{BM3} \cite{zhou2023bootstrap} : This method simplifies the self-supervision task in multimodal recommender systems.
		
		\textbf{MGCN} \cite{yu2023multi} : This method is based on the GCN, which utilizes item information to purify modal features. A behavior-aware fuser is also designed that can adaptively learn different modal features.

            \textbf{DiffMM}  \cite{jiang2024diffmm} :This method is based on diffusion model, which enhances the user's representation by combining cross-modal contrastive learning and modality-aware graph diffusion models for more accurate recommendation results.
  
		\textbf{Freedom} \cite{zhou2023tale} : This method is based on freezing the U-I graph and the I-I graph, and a degree-sensitive edge pruning method is designed to delete possible noisy edges.
  
	\subsubsection{ Evaluation protocols}
	In this subsection, we introduce the evaluation metrics used in the experiments. The evaluation metrics employ in this experiment are Recall@$K$ (R@$K$) and NDCG@$K$ (N@$K$). Recall represents the recall rate, while NDCG stands for Normalized Discounted Cumulative Gain. We set $K = \left\lbrace 10, 20 \right\rbrace $, indicating number of items in final recommendation list.
	\subsubsection{Details}
	This subsection provides a detailed description of the hyperparameter settings of the DiffCL on different datasets. To ensure the fairness of the evaluation, we use MMRec \cite{zhou2023comprehensive} to implement all the comparative baselines and also execute a grid search for hyperparameters of these models to determine the optimal hyperparameter settings. In addition, Adam optimizer is employed to make superior the DiffCL and other models. 
	
	We install learning rate of the DiffCL to 0.001, dropout rate to 0.5, and $\tau$ in graph contrastive learning to 0.4. Besides, the values of $\lambda_{\mathrm{cl}}$, $\lambda_{align}$ and $\lambda_{E}$ are not the same for different datasets. Specifically, we set three different sets of loss weights for the baby, video, and sports datasets, respectively. They are as follows: $\left \{0.1, 0.4, 0.7  \right \} $, $\left \{0.01, 1.0, 1.0   \right \} $ and $\left \{0.7, 0.4, 0.9  \right \} $.
	
	\begin{table*}[ht]
		\centering
		\setlength{\tabcolsep}{1pt}
		\scriptsize
		\setlength{\abovecaptionskip}{5pt}%
		\setlength{\belowcaptionskip}{10pt}%
		\renewcommand{\arraystretch}{1.5} 
		\setlength{\tabcolsep}{7pt} 
		\caption{The Comparison Of Different Baselines And The DiffCL Performance On Three Datasets}
		\vskip -0.1in
		\label{tab:comparison results}
		\begin{tabular}{ccccc|cccc|cccc}
			\hline
			Datasets&  \multicolumn{4}{c}{Baby}&  \multicolumn{4}{c}{Video}&  \multicolumn{4}{c}{Sports}\\\hline\hline
			Model & R@10& R@20& N@10& N@20& R@10& R@20& N@10& N@20& R@10& R@20& N@10& N@20\\\hline
			BPR & 0.0268& 0.0441& 0.0144& 0.0188& 0.0722& 0.1106& 0.0386& 0.0486 & 0.0306 & 0.0465 & 0.0169 & 0.0210\\
			LightGCN & \underline{0.0402}& \underline{0.0644}& \underline{0.0274}& \underline{0.0375}& \underline{0.0873}& \underline{0.1351}& \underline{0.0475}& \underline{0.0599}& \underline{0.0423}& \underline{0.0642}& \underline{0.0229}& \underline{0.0285}\\
			DiffCL &  \textbf{0.0641}&  \textbf{0.0987}& \textbf{0.0343}&  \textbf{0.0433 }&  \textbf{0.1421}&  \textbf{0.2069} &\textbf{0.0804 } & \textbf{0.0974} 
			&  \textbf{0.0754}&  \textbf{0.1095} &\textbf{0.0421} & \textbf{0.0509} \\\hline
			\textbf{Improv.}&  32.50\%&  42.70\%&  14.23\%&  5.60\%&  59.45\%&  50.55\%& 62.11\%&  56.43\% &  64.78\%&  60.59\%&  65.06\%&  64.91\%\\\hline\hline
			VBPR &  0.0397&  0.0665&  0.0210 &  0.0279&  0.1198&   0.1796& 0.0647& 0.0802& 0.0509& 0.0765& 0.0274& 0.0340\\
			MMGCN &  0.0397&  0.0641&  0.0206&  0.0269&  0.0843&   0.1323& 0.0440& 0.0565 & 0.0380& 0.0610 & 0.0206& 0.0266\\
			DualGNN &  0.0518& 0.0820 & 0.0273& 0.0350& {0.1200} &{0.1807}&{0.0656}&{0.0814}& 0.0583& 0.0865& 0.0320 &0.0393\\
			SLMRec & 0.0529& 0.0775&{0.0290}& 0.0353& 0.1187&  0.1767& 0.0642& 0.0792 &0.0663&{0.0990} & {0.0365}& {0.0450}\\
			BM3 &{0.0539}& {0.0848}& 0.0283& {0.0362}& 0.1166& 0.1772& 0.0636& 0.0793& 0.0632& 0.0940 & 0.0346& 0.0426\\
			MGCN &0.0608 &0.0927 & \underline{0.0333} &\underline{0.0415} &\underline{0.1345} & \underline{0.1997}&\underline{0.0740} &\underline{0.0910 }&0.0713 &0.1060 &0.0392 &0.0489 \\
			Freedom & \underline{0.0622}&\underline{0.0948} & 0.0330&0.0414 &0.1226 &0.1858 & 0.0662&0.0827 &\underline{0.0722} &\underline{0.1062} & \underline{0.0394}&\underline{0.0484} \\
			DiffMM &0.0619& 0.0947& 0.0326& 0.0394& --- & ---&---& ---& 0.0683& 0.1019 & 0.0374& 0.0455\\
   
			DiffCL &  \textbf{0.0641}&  \textbf{0.0987}& \textbf{0.0343}&  \textbf{0.0433 }&  \textbf{0.1421}&  \textbf{0.2069} &\textbf{0.0804 } & \textbf{0.0974} 
			&  \textbf{0.0754}&  \textbf{0.1095} &\textbf{0.0421} & \textbf{0.0509} \\

			\textbf{Improv.}& 3.05\%&  4.11\%&  3.93\%&  4.58\%&  5.65\%& 3.60\%& 8.64\%& 7.03\% &  4.43\%&   3.11\%& 6.85\%& 5.16\% \\\hline\hline
		\end{tabular}
		\label{2}
	\end{table*}

\begin{table}[H]
	\centering
	\setlength{\abovecaptionskip}{5pt}%
	\setlength{\belowcaptionskip}{0pt}%
	\setlength{\tabcolsep}{7pt} 
	\renewcommand{\arraystretch}{1.2} 
	\caption{The Performance Comparison Of Different Variants}
	\vskip -0.1in
	\begin{tabular}{ll|ccc}\hline
		\multirow{2}{*}{\textbf{Variants}}& \multirow{2}{*}{\textbf{Metrics}}& \multicolumn{3}{c}{\textbf{Datasets}}\\
		&&Baby& Video&Sports\\ \hline\hline
		\multirow{2}{*}{DiffCL$_{baseline}$}&R@20 &0.0854& 0.1907&0.0956\\
		& N@20& 0.0364& 0.0856&0.0428\\ \hline
		\multirow{2}{*}{DiffCL$_{diff}$} &R@20&{0.0925}& \underline{0.1978}&0.1095\\ 
		&N@20&0.0396& 0.0895& \underline{0.0509}\\\hline
		\multirow{2}{*}{DiffCL$_{align}$} & R@20 & 0.0907&{0.1965}& 0.0960\\
		& N@20 & 0.0392& {0.0893}& 0.0428\\\hline
		\multirow{2}{*}{DiffCL$_{h}$} & R@20 &\underline{{0.0986}}& 0.1921& 0.1099\\
		& N@20 & 0.0430& 0.0872& 0.0494\\\hline
		\multirow{2}{*}{DiffCL$_{diff+align}$} & R@20 &0.0911& 0.1904 & 0.1093\\
		& N@20 & 0.0403&0.0866 &0.0506 \\\hline
		\multirow{2}{*}{DiffCL$_{diff+h}$} & R@20 &\underline{ 0.0986}& 0.1940& 0.1102\\
		& N@20 & 0.0430& 0.0885& 0.0495\\\hline
		\multirow{2}{*}{DiffCL$_{align+h}$} & R@20 & 0.0993& 0.1968& \underline{0.1114}\\
		& N@20 & \underline{0.0432}& \underline{0.0896}&0.0496\\\hline
		\multirow{2}{*}{DiffCL} & R@20 & \textbf{0.0987}& \textbf{0.2069}& \textbf{0.1095}\\ 
		& N@20& \textbf{0.0433}& \textbf{0.0974}& \textbf{0.0509}\\\hline
	\end{tabular}
	\vskip -0.15in
	\label{ablation}
	
\end{table}    
\subsection{Comparative Experiments (RQ1)}
\noindent The experimental results are comprehensively summarized in Table \ref{2}. This table presents the specific performance of DiffCL alongside all comparative models. Bold numbers represent the results for DiffCL, underlined numbers denote the results for the best comparative model, and \textbf{Improv.} denotes the percentage improvement of DiffCL over the best comparative model.\par 
The results of this experiment show that most multimodal recommendation models perform significantly better than general recommendation models. This is due to the ability of multimodal recommendation methods to integrate multimodal information about items, thus enabling better capture of user preference cues. From the experimental results, the DiffCL performs best on the sports dataset compared to the general recommendation model. It improves 64.78\%, 60.59\%, 65.06\% and 64.91\% on the four evaluation metrics R@$10$, R@$20$, N@$10$ and N@$20$ respectively. In contrast, the DiffCL has a smaller boost on the baby dataset, improving only 32.50\%, 42.70\%, 14.23\%, and 5.60\% under each of the four metrics mentioned above. This result shows that on the baby dataset, the multimodal information of the items has little effect on the user's preference, and users may be more concerned about other factors such as the quality and price of the items.\par
\begin{figure}[h]	
	\centering
	\subfigure{\includegraphics[width=0.8\columnwidth]{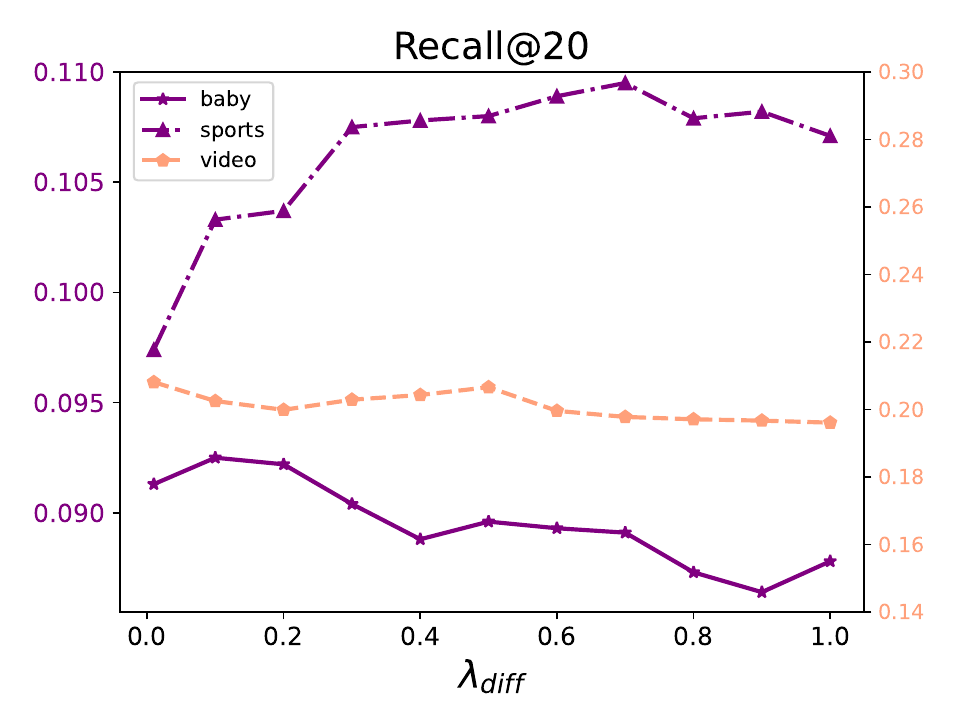}}
	\subfigure{\includegraphics[width=0.8\columnwidth]{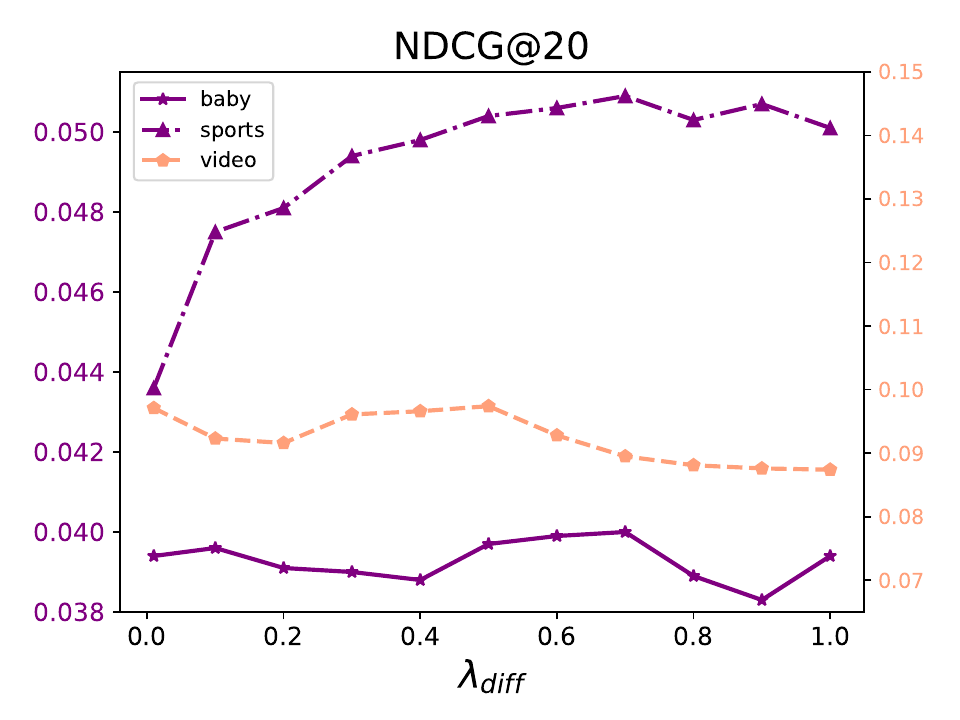}}
	\caption {The performance of the DiffCL under various $\lambda_{diff}$ settings}
	\label{diff}
	\end{figure}
Compared to other multimodal recommendation models, DiffCL consistently outperforms the leading models across all scenarios.
According to the experimental results, DiffCL demonstrates superior performance on the video dataset, with improvements of 5.65\%, 3.60\%, 8.64\%, and 7.03\%  across the four evaluation metrics: R@$10$, R@$20$, N@$10$, and N@$20$, respectively. These results validate the effectiveness of DiffCL in improving recommendation accuracy. Overall, the findings indicate that DiffCL enhances recommendation performance by constructing contrastive views for graph-based contrastive learning through the diffusion model, utilizing the Item-Item (I-I) graph for data augmentation, and employing ID modality-guided inter-modal alignment.

\begin{figure}[h]	
	\centering
	\subfigure{\includegraphics[width=0.8\columnwidth]{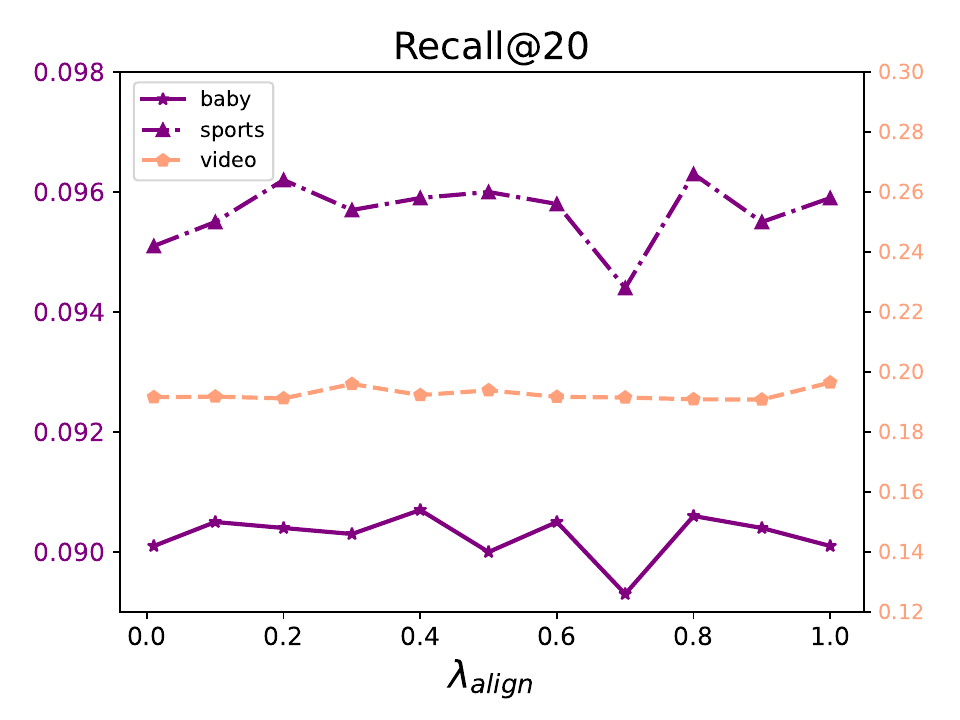}}
	\subfigure{\includegraphics[width=0.8\columnwidth]{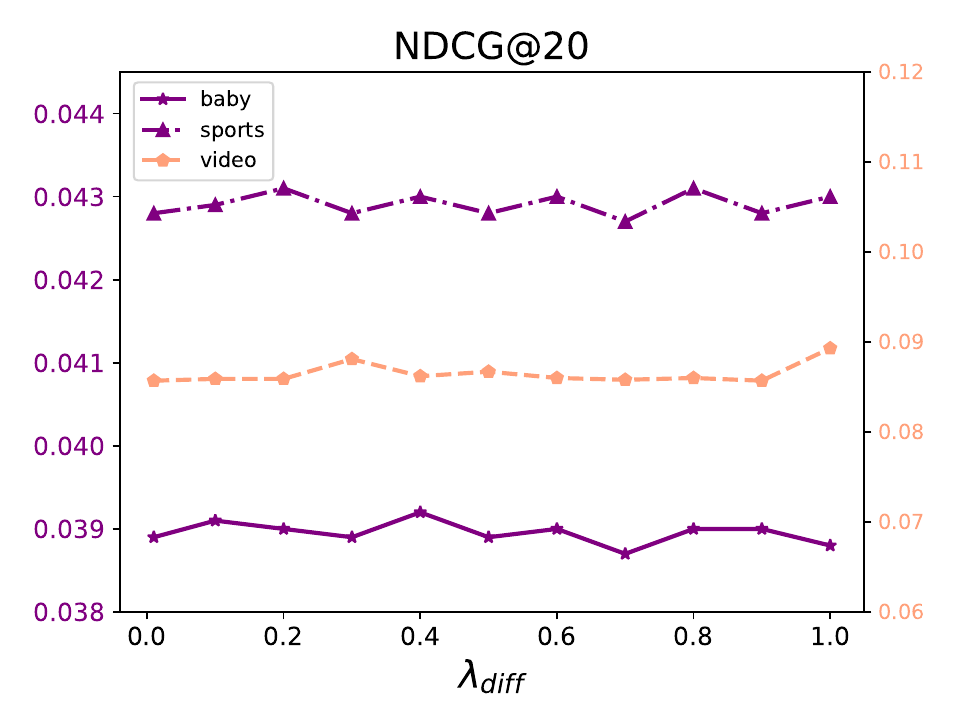}}
	\caption {The performance of the DiffCL under various $\lambda_{align}$ settings}
	\label{ali}
	\end{figure}
\subsection{ Ablation Study (RQ2)}
\noindent In this section, a large number of ablation experiments are performed in order to verify the effectiveness of the various components that make up DiffCL. Specifically, our ablation experiments included the following variants: 
\begin{itemize}
	\item \textbf{DiffCL$_{baseline}$}: Remove all components.
	\item \textbf{DiffCL$_{diff}$}: Retain only the diffusion graph contrastive learning task.
	\item \textbf{DiffCL$_{align}$}: Retain only the ID modal guidance intra-modal semantic alignment  task.
	\item \textbf{DiffCL$_h$}: Retain only the feature enhancement task.
	\item \textbf{DiffCL$_{diff+align}$}: Retain both the diffusion graph contrastive learning task and the ID modal guidance intra-modal semantic alignment task.
	\item \textbf{DiffCL$_{diff+h}$}: Retain both the diffusion graph contrastive learning task and the feature enhancement task.
	\item \textbf{DiffCL$_{align+h}$}: Retain both the ID modal guidance intra-modal semantic alignment task and the feature enhancement task.
\end{itemize}

Table \ref{ablation} shows the results of the final ablation experiments. Bold numbers denote best result and underlined numbers denote sub-optimal results. According to the final results, all components of the DiffCL are valid in improving the performance of the whole system respectively. In addition, the models consisting of the combination of any two components also get better recommendation results compared to the models with a single component.
  
	\begin{figure}[ht]	
	\centering
	\subfigure{\includegraphics[width=0.8\columnwidth]{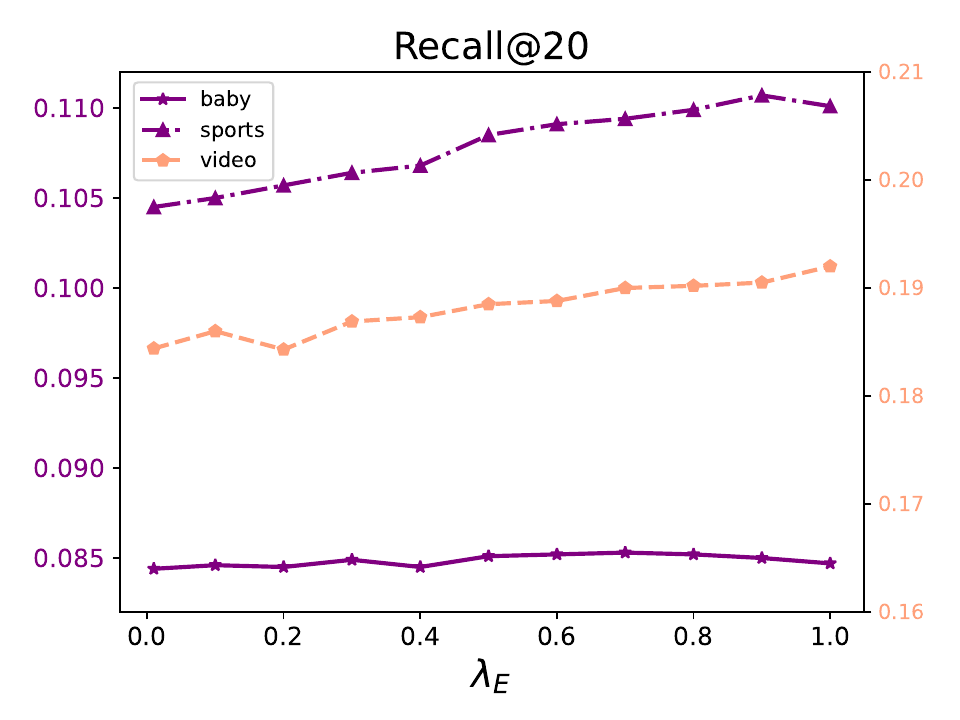}}
	\subfigure{\includegraphics[width=0.8\columnwidth]{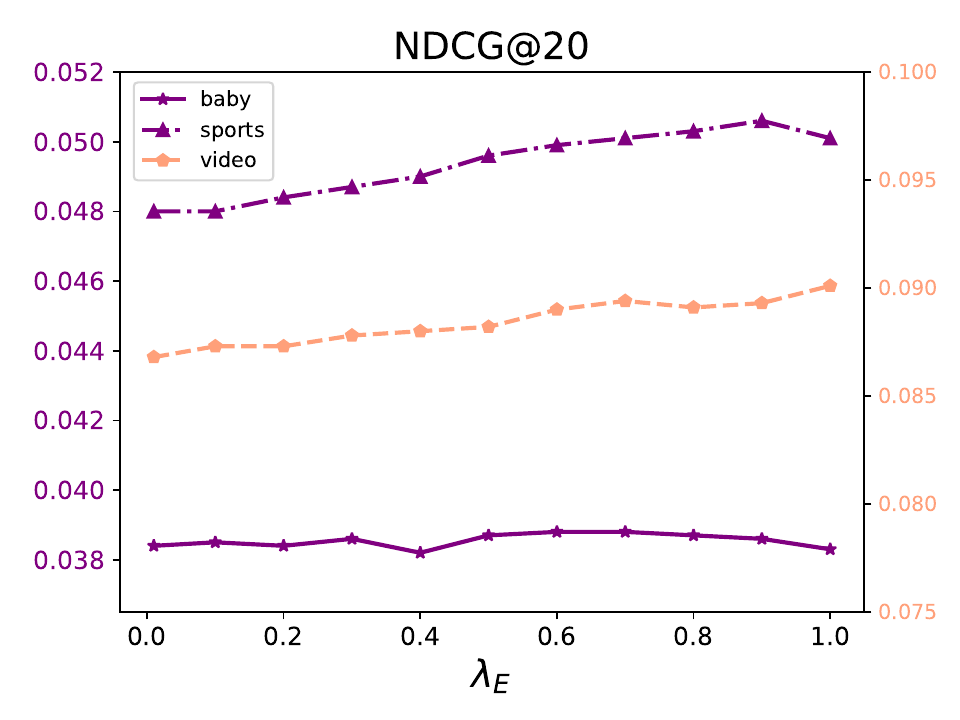}}
	\caption {The performance of the DiffCL under various $\lambda_{E}$ settings}
	\label{zhu}
	\end{figure}
    
\subsection{Hyperparameter Effects (RQ3)}
\noindent 
In this section, we investigate the effect of varying loss weights on the performance of diffusion map contrastive learning and multimodal indirect alignment. Specifically, we conduct a series of hyperparameter experiments to analyze how the values of these weights impact model performance. The loss weights for both the diffusion map contrastive learning task and the multimodal indirect alignment task are set to: $\lambda_{diff}, \lambda_{align}, \lambda_{E} \in \{0.01, 0.1, 0.2, 0.3, 0.4, 0.5, 0.6, 0.7, 0.8, 0.9, 1.0\}$. The experimental results across three datasets are presented in Figures \ref{diff}, \ref{ali}, and \ref{zhu}.The results reveal that the values of the loss weights significantly influence the model's performance. Although the loss weights for the three parameters, $\lambda_{diff}$, $\lambda_{align}$, and $\lambda_{E}$, are set to the same range, the optimal values for each weight vary depending on the dataset and task. Tuning these weights appropriately is crucial for achieving the best performance. The findings emphasize the importance of hyperparameter selection in optimizing the  robustness and accuracy of the Diffcl across different task.
\section{Conclusion}
\noindent In this paper,  a diffusion-based contrastive learning (DiffCL) framework is proposed for multimodal recommendation. This method generates high-quality contrastive views by introducing a diffusion model during the graph contrastive learning stage, effectively addressing the issue of reduced recommendation accuracy caused by noise in self-supervised tasks. Furthermore, it employs stable ID embeddings to guide semantic alignment across different modalities, significantly enhancing the semantic consistency of items. To comprehensively evaluate the performance of the DiffCL , we conduct a series of experiments on multiple real-world datasets and compare it with various recommendation models. The experimental results demonstrate the effectiveness of each component of the DiffCL and its superiority in recommendation performance.

In future research, we aim to optimize the integration of diffusion models within recommender systems, extending their application beyond specific stages of the recommendation process. By leveraging the powerful generative capabilities of the diffusion model, we intend to perform data augmentation from multiple perspectives to achieve superior recommendation outcomes.

\bibliographystyle{IEEEtran}
\bibliography{ref}

\end{document}